\documentclass[prl,aps,showpacs,superscriptaddress,twocolumn]{revtex4}
\usepackage{amsmath,amssymb,amsfonts,graphicx}
\newcommand{\ket}[1]{\displaystyle{|#1\rangle}}

\newcommand{\M}{{\scriptscriptstyle M}}
\newcommand{\A}{{\rm A}}
\newcommand{\E}{{\scriptscriptstyle E}}
\newcommand{\R}{{\scriptscriptstyle R}}
\newcommand{\0}{{\scriptscriptstyle 0}}
\newcommand{\NM}{{\scriptscriptstyle N\!M}}
\newcommand{\NE}{{\scriptscriptstyle \!N\!E}}
\newcommand{\ac}{a^{\dag}}

\begin{document}
\title{Continuous variable quantum key distribution in non-Markovian channels}
\author{Ruggero Vasile}\email{ruggero.vasile@utu.fi}
\affiliation{Turku Centre for Quantum Physics, Department of Physics
and Astronomy, University of Turku, FI-20014 Turun yliopisto, Finland}
\author{Stefano Olivares}
\affiliation{CNISM UdR Milano Universit\`a, I-20133 Milano, Italy}
\author{Matteo G. A. Paris}
\affiliation{Dipartimento di Fisica, Universit\`a degli Studi di Milano,
I-20133 Milano, Italy}
\author{Sabrina Maniscalco}
\affiliation{Turku Centre for Quantum Physics, Department of Physics
and Astronomy, University of Turku, FI-20014 Turun yliopisto, Finland}
\begin{abstract}
We address continuous variable quantum key distribution (QKD) in non-Markovian lossy
channels and show how the non-Markovian
features may be exploited to enhance security and/or to detect the
presence and the position of an eavesdropper along the transmission
line.  In particular, we suggest a coherent states QKD protocol which is
secure against individual attacks for arbitrarily low values of the
overall transmission line. Our scheme relies on specific non-Markovian
properties, and cannot be implemented in ordinary Markovian channels
characterized by uniform losses.
\end{abstract}
\date{\today}
\maketitle Quantum key distribution (QKD) is a fundamental area of
quantum technology \cite{Gis02}. The aim of any QKD protocol is to allow two
parties, the sender Alice and the receiver Bob, to exchange a secret
key using quantum and/or classical channels, avoiding a possible
eavesdropper (Eve) to acquire information on the key.  Discrete
variable QKD protocols are based on transmission and measurements of
single or entangled photons states, and therefore they are limited
by the efficiency of single photon generation and detection.  
On the contrary, continuous variable (CV) QKD \cite{cer07}
is a potentially high-bit-rate technique for at least two reasons.
One the one hand, the key is encoded into continuous-spectrum quantum 
observables as the quadrature components of a light field, and thus 
the number of bits per pulse can be high. On the other hand, it 
employs homodyne detection technique based on standard photodiodes, 
which are much faster than the avalanche photodiodes used in 
photon-counting discrete QKD schemes. 
Different proposals for CV QKD have been put forward, either based on
single-mode coherent \cite{GroGra02,NatGroGra} and squeezed
\cite{Hil00,Ral00,Got01,Cer01,Ben01,Sil02,gp09} signals or EPR
correlated beams \cite{Rei00,Nat02}.  Experimental demonstrations
have been reported for coherent state
\cite{GroGra02,NatGroGra,Hir03,Qi07,Sym07}, squeezed \cite{Sil02}
and EPR beams \cite{Nat02} based protocols, and unconditional
security proofs have been also given \cite{lev09}. In the coherent
state protocol, which is the most interesting for practical
applications,  Alice encodes a key into amplitudes of pure coherent
states and sends them through a quantum channel to Bob, who randomly
chooses a coding quadrature basis in which to measure via homodyne
detection. Binary data is extracted from the homodyne sample using
the bit-slice reconciliation method and privacy amplification
\cite{VA04}.
\par
The unavoidable losses occurring along the channel must be taken
into account for a realistic description of any QKD protocol. In
fact, it has been shown that losses can be exploited by
eavesdroppers to hide themselves \cite{GroGra02}. Security of the
protocol is then defined through the equivalent noise referred to
the input, i.e. on ensuring that the information of Bob about the
key to be higher than the one acquired by Eve. Lossy
channels considered so far are Markovian, i.e. characterized by a
damping rate which is constant along the transmission line.
This is usually an approximation and in practice channels may show
non-Markovian losses, i.e. a damping rate which is not uniform along
the transmission line, being dependent on the spectral structure of
the environment coupled to the propagating mode \cite{Cor01,Cor06}. Moreover, the increasing success of reservoir engineering techniques paves the way to the realization of optical channels in which the losses due to the interaction with the environment can be appropriately manipulated. Recently, e.g.,  non-Markovian signatures in semiconductor quantum wires have been experimentally observed \cite{Marq09}.
\par
In this Letter we address for the first time the effects of
non-Markovian channel losses on the performances of a CV QKD
protocol. In particular, we focus on the coherent state protocol and
show how the non-Markovian features may be exploited to enhance
security, i.e. to reduce the information available for Eve and/or to detect her presence and position along the
transmission line. In our scheme, a suitable engineering of the
channel decay rate allows us to obtain secure QKD for arbitrarily
low values of the overall transmission line. We also show that the
same result cannot be obtained with ordinary Markovian channels
characterized by uniform losses. In the following, we briefly review
the coherent state protocol in a Markovian channel and describe the
eavesdropping strategy. We then introduce a relevant class of
non-Markovian channel and generalize the protocol to this case.
Finally, we describe in detail our proposal to enhance security and
show how it is possible to detect the eavesdropper by a
post-communication comparison of part of the data sent by Alice to
Bob. We also discuss how to optimize the decay properties of the
channel for a specific type of structured environment.
\par
\emph{QKD using coherent states} --- In QKD with coherent states
\cite{GroGra02}, Alice draws pairs of indepedent real random numbers
($x_\A$, $p_\A$) from two Gaussian distributions with zero mean and
the same variance, and then generates the coherent states
$\ket{\alpha_\A}=\ket{x_\A+i p_\A}$, which are finally sent to Bob
through a quantum channel. The propagation along the channel is in
general noisy and losses are described as the interaction of the
light mode with an environment made of a zero temperature ensemble
of independent harmonic oscillators under the Born-Markov
approximation \cite{Breuer}. The evolution is thus governed by a
Master equation in the Lindblad form
$\dot{\varrho}=\gamma(2a\varrho\ac-\ac a\varrho-\varrho\ac a)$. The
state sent by Alice evolves as
$\ket{\alpha_\A}\rightarrow\ket{\alpha_\A e^{-\gamma
    t}}\equiv\ket{\alpha_\A\sqrt{\eta_\M}}$, where $\eta_\M\equiv
\eta_\M (t)=e^{-2\gamma t}$ is the channel transmission. We denote
by $\tau\equiv L$ (hitherto $c=1$) the total transmission
time/channel length.  After the propagation Bob receives the damped
states and arbitrarily decides to measure one of two orthogonal
quadratures. Since the key is encoded in the mean value of the
signal sent by Alice, he needs to rescale the measured
observables by an amount equal to $\eta_\M (\tau)^{-1/2}$, thus also
amplifying the noise.
\par
\emph{Eavesdropping strategy} --- Due to the very nature of the
protocol, the best passive attack Eve may perform is the use of
an optimal asymmetric $1\rightarrow 2$ cloning machine
\cite{Clo,OptEve,NaKoa}. This process can be modeled with Eve
intercepting the signal with a beam splitter of transmissivity
$\eta_\E$ at position $L_\E=t_\E$ along the line. We assume Eve
knows the features of the quantum channel (the lenght
$\tau$ and the loss rate $\gamma$) and that she can tune with
arbitrary precision both the value of the transmissivity $\eta_\E$
and the attack time $t_\E$. The reflected part of the beam is stored
by Eve whereas the other part is sent to Bob through a lossless
channel. Under these conditions, the best eavesdropping strategy is
to attack immediately ($t_\E=0$) with a beam splitter of
transmissivity $\eta_\E=\eta_\M(\tau)$ \cite{GroGra02} equal to the
overall transmissivity of the channel. In this way, Eve is
introducing the same amount of losses as the overall line: Bob will
receive the same state as in the absence of any attack and the 
eavesdropper is not detectable. If however $\eta_\M(\tau)\geq
1/2$ then, even if not detected, Eve cannot achieve the same
information as Bob about the secret key and the protocol is secure
\cite{GroGra02}.
\par
\emph{Non-Markovian channels} --- Markovian evolutions are
approximate dynamical models for channel losses and more realistic
situations can be described with Master equations derived without
the Markov assumption. As for example, including the non-resonant
coupling to phonons in the description of propagation in fused
silica fibers, leads to delayed nonlinearity due to the non-Markovian
phonon bath in addition to spontaneous and thermal noise
\cite{Cor06}. In the following we consider the non-Markovian master
equation (NME) $\dot{\varrho}=\gamma(t)(2a\rho\ac-\ac a\rho-\rho\ac
a)$, which corresponds to a  model in which the light mode
interacts weakly with a structured bosonic reservoir at zero temperature. The functional form of the
coefficient $\gamma(t)$ depends on the spectral structure of the
environment in which the system is embedded. In the weak coupling
regime and for times larger than the typical reservoir correlation
time scale $\tau_\R$, the coefficient tends to the Markovian
constant value, i.e., $\gamma(t)\rightarrow\gamma_\M$. By changing
the reservoir spectral properties one may engineer the functional
form of $\gamma(t)$ as well as modify the value of $\tau_\R$. It is worth noticing that the key feature in our scheme is the inhomogeneity in the rate of loss $\gamma$. This can also be achieved by a suitably engineered position-dependent coupling to the reservoir along the optical channel, since $\gamma (x)=\gamma (c t)$. In a
non-Markovian channel an initial coherent state evolves as
$\ket{\alpha_\A}\rightarrow\ket{\alpha_\A
e^{-\Gamma(t)/2}}=\ket{\alpha_\A\sqrt{\eta_{\NM}}}$ where
$\eta_{\NM}(t)=e^{-\Gamma(t)}$ is the channel transmissivity, with
$\Gamma(t)=2\int_0^t\gamma(s) ds$. Because of the time dependence of
the coefficient $\gamma(t)$, we have in general
$\Gamma(t)\not\propto t$, i.e. the damping is not uniform as in the
Markovian case. The eavesdropping strategy described above works in
the same way if we let Eve  known the analytic form of the decay
rate $\gamma(t)$. In this case, the best strategy is still to attack
at the beginning of the channel and to choose properly the beam
splitter transmissivity to have $\eta_\E=\eta_{\NM}(\tau)$. In this
way her presence is still non detectable and the results about the
security of the channel reported in \cite{GroGra02,NatGroGra} still
hold. On the other hand, the time-dependent losses may be exploited
to detect the presence of the eavesdropper and, in some cases, its
position along the line.
\par
\begin{figure}[t!]
\centerline{\includegraphics[width=0.8\columnwidth]{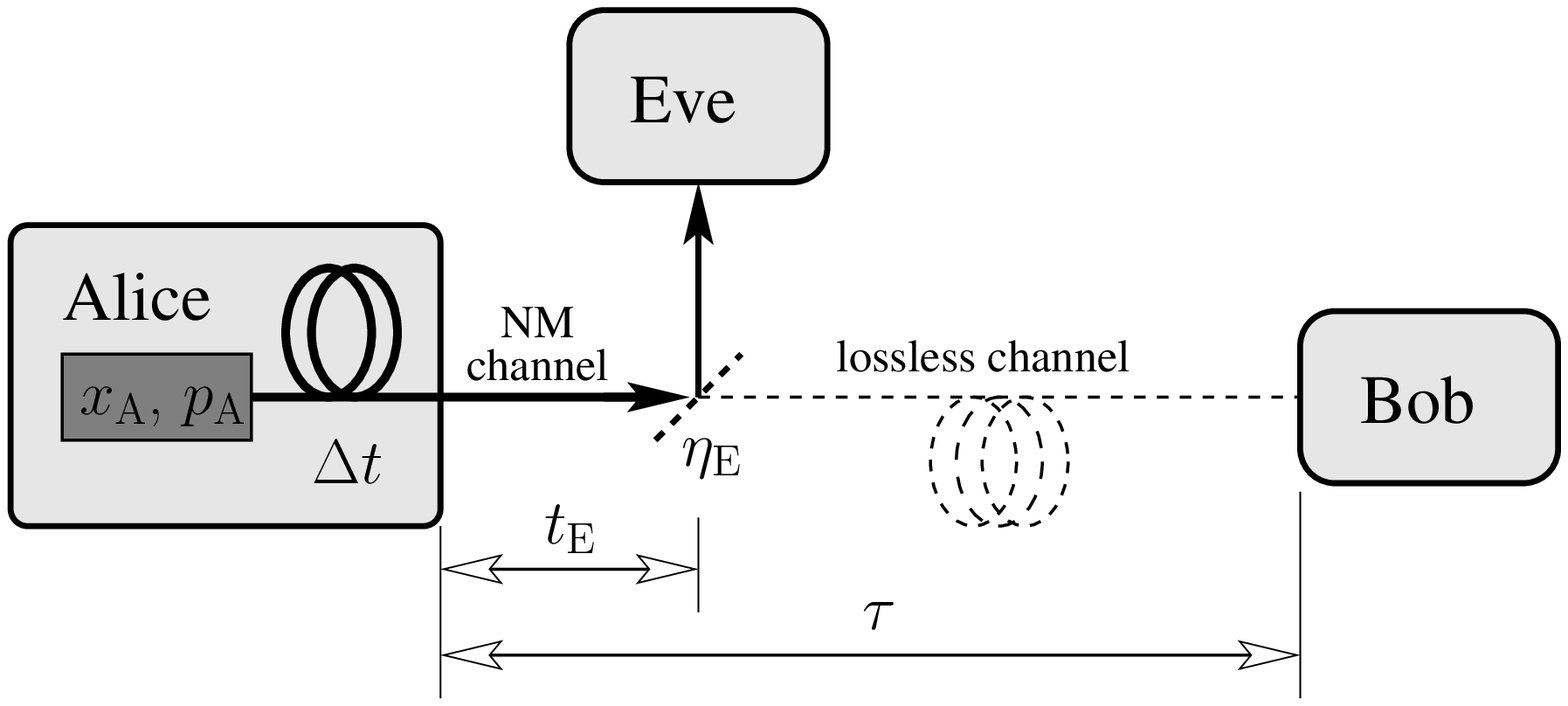}}
\caption{Diagram of the QKD protocol in NM channel with
the relevant elements of Eve's attack scheme.
The solid channel line refers to the NM channel, the dashed one
refers to the lossless channel used by Eve. Without the eavesdropper,
the channel is non-Markovian throughout its whole length.
\label{f:sch}}
\end{figure}
\emph{Eavesdropping detection in NM channels} --- One of the main
assumptions in QKD is that everything Alice communicates to Bob
using public channels is also known by Eve. This means that, in
order to detect a possible eavesdropper, Alice needs to perform
independently a certain operation during the transmission, leading
to different results at Bob side when Eve is present or not. In our
protocol, Alice still encodes the key into coherent signals, but now
she can act on the channel length by adding a delay $\Delta t$ at
the first stage of the signal propagation, as depicted in
Fig.~\ref{f:sch}. For the sake of simplicity, though we have only
one physical channel, we will refer to the two possible choices as
two channels with the same time-dependent loss rate $\gamma(t)$ but
different length. The key signal is always sent through the ordinary
channel of length $\tau$, but now Alice may also send a reference
coherent state $\ket{\alpha_\0}$ by choosing randomly, with the same
probability, between the ordinary and the longer channel of length
$\tau+\Delta t$.
\par
Let us start by considering the situation of clean channel (no
eavesdropper) and focus to the results of the quadrature
measurements for the reference state. Bob is receiving fifty percent
of the time the state $\ket{\alpha_\0 e^{-\Gamma(\tau)/2}}$ when it
is sent through the ordinary line whereas the rest of the copies
evolve into $\ket{\alpha_\0 e^{-\Gamma(\tau+\Delta t)/2}}$. Since
Bob is not aware of which channel has been chosen by Alice, his
quadrature measurements must be independent from this choice.
Therefore he measures quadratures scaled according to the total
ordinary channel losses $e^{-\Gamma(\tau)/2}$. After the completion
of the session Alice informs Bob about which channel each reference
state has been sent through. Bob can now distinguish among the two
sets of states and study the statistics of the two measurement
distributions. It is easy to show that these distributions are
Gaussian with same width but they differ in the mean value by an
amount
\begin{align}
\delta x_{\NE}&=|\alpha_\0(1-e^{-(\Gamma(\tau+\Delta t)-
\Gamma(\tau))/2})|\simeq|\alpha_\0\gamma(\tau)\Delta t|\,,\label{MeanNoEVE}
\end{align}
where we assumed that $\Delta t$ is small compared to the variation
of $\gamma(t)$. Moreover the difference in mean value increases as
the amplitude $\alpha_\0$ increases.
\par
The situation changes when Eve is attacking the line. Because she
also cannot distinguish between the reference signals and the key
ones, as well as between the choice of channel by Alice, she has to
treat every state on the same foot. She then keeps unchanged the
attacking time $t_\E$ and the beam splitter transmissivity
$\sqrt{\eta_\E}$. If she attacks at $t_\E$ the transmissivity must
be chosen in a way that
$e^{-\Gamma(t_\E)/2}\sqrt{\eta_\E}=e^{-\Gamma(\tau)/2}$ so to be
undetectable when the ordinary channel is used. If Alice uses the
longer channel Eve's attack time is forcefully shifted to
$t_\E+\Delta t$. The same calculation as before for the difference
in the mean values of the quadrature measurement distribution at Bob
side leads to
\begin{align}\label{MeanEVE}
\delta x_\E&=|\alpha_\0(1-e^{-(\Gamma(t_\E+\Delta t)-\Gamma(t_\E))/2})|\simeq|\alpha_\0\gamma(t_\E)\Delta t|
\end{align}
Because of the time dependence of the loss rate $\gamma(t)$ the
quantities in Eqs. \eqref{MeanNoEVE} and \eqref{MeanEVE} are in general
different. Therefore, if after the communication Alice and Bob perform a
check of the mean values of the distributions using a public channel,
they are able to detect the presence of Eve whenever
$\gamma(t_\E)\neq\gamma(\tau)$. This condition cannot be satisfied in a
Markovian channel.
\par
Certain types of non-Markovian
channels also introduce thermal noise \cite{LindSab}. Since the added noise is time
dependent, besides the mean values also the widths of the
distributions at Bob side are different in the
presence or absence of an eavesdropper. In turn, this may be exploited
to further enhance security via checking the sample variances.
\par
\emph{Discussion} --- Our proposal is based on the fact that, in a
non-Markovian channel, $\Gamma(t+s)\neq\Gamma(t)+\Gamma(s)$ for
generic $t,s\geq0$. The lack of the semigroup property immediately
implies that the integral expression $\int_{t}^{t+\Delta
t}\gamma(s)ds$ for fixed $\Delta t$ is not a constant function of
$t$. Therefore, the two quantities in Eqs. \eqref{MeanNoEVE} and
\eqref{MeanEVE} do not coincide and the only way Eve can avoid
detection is to attack the channel when
$\gamma(t_\E)\simeq\gamma(\tau)$. As a consequence, it is crucial to engineer appropriately the environment surrounding the channel to obtain the
desired decay properties. As a concrete example we consider here the
decay rate evaluated for an Ohmic reservoir with Lorentz-Drude
cutoff \cite{Weiss}, i.e. 
$\gamma(t)=\gamma_\M\bigl[1-e^{-\omega_c
t}\cos\omega_\0t-(\omega_c / \omega_\0)e^{-\omega_c
t}\sin\omega_\0t\bigl]$,
$\omega_\0$ being the mode frequency, $\omega_c$ the
cut-off frequency of the environment spectrum, and $\gamma_\M$ the
asymptotic decay rate. The reservoir correlation time is here
identified with $\tau_\R=\omega_c^{-1}$.
\begin{figure}[t!]
\centerline{\includegraphics[width=0.8\columnwidth]{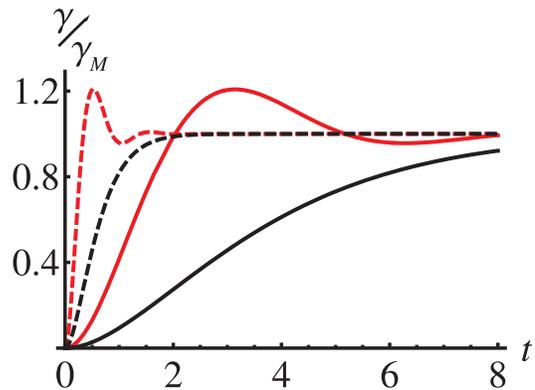}}
\caption{(Colors online) Normalized decay rates
(see text) as a function of time. Red
lines correspond to
$\omega_c/\omega_0=0.5$ and black lines to $\omega_c/\omega_0=3$.
$\omega_c=0.5$ for solid lines, and $\omega_c=3$ for dashed lines.}
\label{f:f2dec}
\end{figure}
\par
In Fig. \ref{f:f2dec} we show the behavior of the normalized decay
rate $\gamma(t)/\gamma_\M$ for different values of light and cut-off
frequencies $\omega_\0$ and $\omega_c$. The red lines are evaluated
for the same ratio $\omega_c/\omega_0=0.5$ and differ for the value
of $\omega_c=0.5$ (solid line) and $\omega_c=3$ (dashed line). The
black lines instead correspond to $\omega_c/\omega_0=5$ and
$\omega_c=0.5$ (solid line), $\omega_c=3$ (dashed line). Decay rates
of this second class (regime $\omega_c>\omega_0$) are exactly what
is needed for our scheme to work, because the relation
$\gamma(t_\E)<\gamma(\tau)$ holds for any allowed value of the
attack time $t_\E$ and Eve has in principle no way to hide herself
from the security protocol. Moreover, since for
$\omega_c>\omega_\0$ the function $\gamma(t)$ is invertible, Alice
and Bob can also find the exact position of the eavesdropper along
the line. On the other hand decay functions represented by the red
lines present oscillations before relaxing to the stationary
value and are not invertible. In this case Eve may find several
places at the beginning of the line where she can perform the attack
while avoiding detection, i.e. when $\gamma(t_E)=\gamma_M$.
\par
Our detection method relies on checking whether the distributions of
homodyne data from the two channels of different lengths are shifted
each other by an amount $\delta x_\NE$ rather than $\delta x_\E$.
The ability of detecting an eavesdropper thus depends on the precision
and the resolution of quadrature measurements made by Bob. A finite
precision implies that Bob could not be able to discriminate the
results when $|\alpha_0 \Delta t
[\gamma(t_\E^*)-\gamma(\tau)]|<\epsilon$, with $\epsilon$  a
threshold depending on the precision. In practice, this means that
whenever Eve places the attack at $t_\E>t_\E^*$ she is not revealed
by our method.  According to Ref. \cite{GroGra02} the channel is then
secure when $\Gamma (t_\E)\geq - \log 2 \eta_\NM$, $\eta_\NM = \exp
\{-\Gamma (\tau)\}$ being the overall transmission of the
non-Markovian channel. In other words, security is ensured if the
overall transmission is larger than $\eta_\NM \geq \frac12
\exp\{-\Gamma (t_\E^*)\} \equiv \eta_{th}$. For any given $\epsilon$
we can make $t_\E^*$ in principle arbitrarily close to $\tau$ by a
suitable engineering of the environment spectrum. In turn, this
allows one to decrease the threshold and make the channel secure for 
arbitrarily low values of the overall transmission $\eta_\NM$.
Notice also that, being $t_\E$ of the order of the reservoir correlation time scale
$\tau_\R$, if $\tau_\R\ll\tau$ then the amount of
losses accumulated before the attack are negligible compared to the
overall losses and the advantage given by our protocol cannot be
appreciated. In order to obtain a consistent improvement we need 
$\tau_\R$ to be of the order of the total transmission time
$\tau$. 
\par
\emph{Conclusion} --- We have analyzed continuous variable QKD with
coherent states in the presence of non-Markovian effects along the
transmission line and suggested a novel method to improve security
based on the non uniform time dependence of the losses. Our method
ensures security for arbitrarily low transmissivity of the channel
and allows one to detect the presence and the position of the
eavesdropper upon both a suitable engineering of the channel decay
properties and the use of an additional reference coherent signal.
The eavesdropper can manage to hide her presence by reducing the
extracted amount of information, but the legitimate users can reduce
to zero her information by tuning the reservoir correlation time.
Our detection scheme is based on a specific non-Markovian property,
and it cannot be implemented in ordinary Markovian channel
characterized by uniform losses. Besides, since it is based on channel 
properties rather than on specific features of the distribution scheme, 
we foresee its application to other CV QKD protocols, as those based on 
squeezed or entangled states.
\par
This work has been supported by the Turku Collegium of Science and
Medicine, the Finnish Cultural Foundation (Science Workshop on
Entanglement), the Magnus Ehrnrooth Foundation, the Emil Aaltonen
Foundation, and partially supported by the CNR-CNISM agreement. 


\begin{thebibliography}{99}
\bibitem{Gis02} N. Gisin, G. Ribordy, W. Tittel, H. Zbinden, 
Rev. Mod. Phys. {\bf 74}, 145 (2002).
\bibitem{cer07} N. J. Cerf, Ph. Grangier, J. Opt. Soc. Am. B {\bf 24},
324 (2007).
\bibitem{GroGra02} F. Grosshans, Ph. Grangier, Phys. Rev. Lett.
\textbf{88}, 057902 (2002).
\bibitem{NatGroGra} F. Grosshans, G. Van Assche, J. Wenger,
R. Brouri, N.J. Cerf, and Ph. Grangier, Nature (London)
\textbf{421}, 238 (2003).
\bibitem{Hil00} M. Hillery, Phys. Rev. A \textbf{61}, 022309 (2000).
\bibitem{Ral00} T. C. Ralph, Phys. Rev. A \textbf{61}, 010303 (1999);
{\bf 62} 062306 (2000).
\bibitem{Got01} D. Gottesman, J. Preskill, Phys. Rev. A {\bf 63}, 022309
(2001).
\bibitem{Cer01} N. J. Cerf, M. Levy, G. Van Assche, Phys. Rev. A {\bf
63}, 052311 (2001).
\bibitem{Ben01} K. Bencheikh, Th. Symul, A. Jankovic, J. A. Levenson,
J. Mod. Opt. {\bf 48}, 19031920 (2001).
\bibitem{Sil02} Ch. Silberhorn, T.C. Ralph, N. Lutkenhaus, G. Leuchs,
Phys. Rev. Lett. {\bf 89}, 167901 (2002).
\bibitem{gp09} R. Garcia-Patron, N. J. Cerf, Phys. Rev. Lett. {\bf 102},
130501 (2009).
\bibitem{Rei00} M. D. Reid, Phys. Rev. A \textbf{62}, 062308 (2000).
\bibitem{Nat02} Ch. Silberhorn, N. Korolkova, G. Leuchs, Phys. Rev.
Lett. {\bf 88}, 167902 (2002).
\bibitem{Hir03} T. Hirano, H. Yamanaka, M. Ashikaga, T. Konishi, R. Namiki,
Phys. Rev. A {\bf 68}, 042331 (2003).
\bibitem{Qi07} B. Qi, L. Huang, L. Qian, H. Lo, Phys. Rev. A {\bf 76}, 052323 (2007).
\bibitem{Sym07} T. Symul, D. J. Alton, S. M. Assad, A. M. Lance, C.
Weedbrook, T. C. Ralph, P. K. Lam, Phys. Rev. A {\bf 76}, 030303 (2007).
\bibitem{lev09} A. Leverrier, E. Karpov, Ph. Grangier, N. J. Cerf,
New J. Phys. {\bf 11}, 115009 (2009).
\bibitem{VA04} G. Van Assche, J. Cardinal , N. J. Cerf, IEEE Trans.
Inform. Th. {\bf 50}, 394 (2004).
\bibitem{Cor01}
P. D. Drummond and J. F. Corney,  J. Opt. Soc. Am. B {\bf 18}, 139 (2001).
\bibitem{Cor06}
J. F. Corney, P. D. Drummond, J. Heersink, V. Josse, G. Leuchs,
U. L. Andersen, Phys. Rev. Lett. {\bf 97}, 023606 (2006).
\bibitem{Marq09}
V. Lopez-Richard, J. C. Gonzalez, F. M. Matinaga, C. Trallero-Giner, 
E. Ribeiro, M. Rebello Sousa Dias, L. Villegas-Lelovsky, and G. E. Marques, 
Nano Lett. {\bf 9}, 3129 (2009).
\bibitem{Breuer} H.\,-\,P. Breuer and F. Petruccione, \emph{The Theory of Open
Quantum Systems} (Oxford University Press, 2002).
\bibitem{Clo}
S. L. Braunstein, V. Buzek and M. Hillery, Phys. Rev. A {\bf 63},
052313 (2001).
\bibitem{OptEve} F.~Grosshans and Ph.~Grangier, Phys. Rev. A {\bf 64},
  010301 (2001).
\bibitem{NaKoa}
R. Namiki, M. Koashi, and N. Imoto, Phys. Rev. A \textbf{73}, 032302
(2006).
\bibitem{Cerf01}
N.J. Cerf, M. Levy, and G. Van Assche, Phys. Rev. A \textbf{63},
052311 (2001).
\bibitem{LindSab}
S. Maniscalco, J. Piilo, F. Intravaia, F. Petruccione, and A. Messina
Phys. Rev. A \textbf{70}, 032113 (2004). 
\bibitem{Weiss}
U. Weiss, \emph{Quantum Dissipative Systems, 2nd edition} (World
Scientific Publishing, Singapore, 1999).
\end{thebibliography}
\end{document}